\documentclass[preprint,unsortedaddress,superscriptaddress,showpacs,preprintnumbers,amsmath,amssymb]{revtex4}

\usepackage{graphicx}
\usepackage{dcolumn}
\usepackage{amsmath}
\usepackage{url}
\def\xmax   {X_{max}} 
 
\def\xmean  {<\hspace{-0.2cm}X_{max}\hspace{-0.2cm}>}

\def\loge   {\log{E}}

\begin{document}


\title{Indications of Proton-Dominated Cosmic Ray Composition above 1.6~EeV}

\author{R.U.~Abbasi}
\affiliation{University of Utah, Department of Physics, Salt Lake
  City, UT, USA} 

\author{T.~Abu-Zayyad}
\affiliation{University of Utah, Department of Physics, Salt Lake
  City, UT, USA} 

\author{M.~Al-Seady}
\affiliation{University of Utah, Department of Physics, Salt Lake
  City, UT, USA} 

\author{M.~Allen}
\affiliation{University of Utah, Department of Physics, Salt Lake
  City, UT, USA} 

\author{J.F.~Amman}
\affiliation{Los Alamos National Laboratory, Los Alamos, NM, USA}

\author{R.J.~Anderson}
\affiliation{University of Utah, Department of Physics, Salt Lake
  City, UT, USA} 

\author{G.~Archbold}
\affiliation{University of Utah, Department of Physics, Salt Lake
  City, UT, USA} 

\author{K.~Belov}
\affiliation{University of Utah, Department of Physics, Salt Lake
  City, UT, USA} 

\author{J.W.~Belz}
\email[Corresponding author: ]{belz@physics.utah.edu}
\affiliation{University of Utah, Department of Physics, Salt Lake
  City, UT, USA}

\author{D.R.~Bergman}
\affiliation{University of Utah, Department of Physics, Salt Lake
  City, UT, USA}
\affiliation{Rutgers University --- The State University of New
  Jersey, Department of Physics and Astronomy, Piscataway, NJ, USA}

\author{S.A.~Blake}
\affiliation{University of Utah, Department of Physics, Salt Lake
  City, UT, USA}

\author{O.A.~Brusova}
\affiliation{University of Utah, Department of Physics, Salt Lake
  City, UT, USA}

\author{G.W.~Burt}
\affiliation{University of Utah, Department of Physics, Salt Lake
  City, UT, USA}

\author{C.~Cannon}
\affiliation{University of Utah, Department of Physics, Salt Lake
  City, UT, USA}

\author{Z.~Cao}
\affiliation{University of Utah, Department of Physics, Salt Lake
  City, UT, USA}

\author{W.~Deng}
\affiliation{University of Utah, Department of Physics, Salt Lake
  City, UT, USA}

\author{Y.~Fedorova}
\affiliation{University of Utah, Department of Physics, Salt Lake
  City, UT, USA}

\author{C.B.~Finley}
\affiliation{Columbia University,  Department of Physics and Nevis
  Laboratory, New York, New York, USA}

\author{R.C.~Gray}
\affiliation{University of Utah, Department of Physics, Salt Lake
  City, UT, USA}

\author{W.F.~Hanlon}
\affiliation{University of Utah, Department of Physics, Salt Lake
  City, UT, USA}

\author{C.M.~Hoffman}
\affiliation{Los Alamos National Laboratory, Los Alamos, NM, USA} 

\author{M.H.~Holzscheiter}
\affiliation{Los Alamos National Laboratory, Los Alamos, NM, USA}

\author{G.~Hughes}
\affiliation{Rutgers University --- The State University of New
  Jersey, Department of Physics and Astronomy, Piscataway, NJ, USA}

\author{P.~H\"{u}ntemeyer}
\affiliation{University of Utah, Department of Physics, Salt Lake
  City, UT, USA}

\author{B.F~Jones}
\affiliation{University of Utah, Department of Physics, Salt Lake
  City, UT, USA}

\author{C.C.H.~Jui}
\affiliation{University of Utah, Department of Physics, Salt Lake
  City, UT, USA}

\author{K.~Kim}
\affiliation{University of Utah, Department of Physics, Salt Lake
  City, UT, USA}

\author{M.A.~Kirn}
\affiliation{Montana State University, Department of Physics, Bozeman,
  MT , USA}

\author{E.C.~Loh}
\affiliation{University of Utah, Department of Physics, Salt Lake
  City, UT, USA}

\author{J. Liu}
\affiliation{Institute of High Energy Physics, Beijing, China}

\author{J.P.~Lundquist}
\affiliation{University of Utah, Department of Physics, Salt Lake
  City, UT, USA} 

\author{M.M.~Maestas}
\affiliation{University of Utah, Department of Physics, Salt Lake
  City, UT, USA}

\author{N.~Manago}
\affiliation{University of Tokyo, Institute for Cosmic Ray Research,
  Kashiwa, Japan}

\author{L.J.~Marek}
\affiliation{Los Alamos National Laboratory, Los Alamos, NM, USA} 

\author{K.~Martens}
\affiliation{University of Utah, Department of Physics, Salt Lake
  City, UT, USA}

\author{J.A.J.~Matthews}
\affiliation{University of New Mexico, Department of Physics and
  Astronomy, Albuquerque, NM, USA }

\author{J.N.~Matthews}
\affiliation{University of Utah, Department of Physics, Salt Lake
  City, UT, USA}

\author{S.A.~Moore}
\affiliation{University of Utah, Department of Physics, Salt Lake
  City, UT, USA}

\author{A.~O'Neill}
\affiliation{Columbia University,  Department of Physics and Nevis
  Laboratory, New York, New York, USA}

\author{C.A.~Painter}
\affiliation{Los Alamos National Laboratory, Los Alamos, NM, USA} 

\author{L.~Perera}
\affiliation{Rutgers University --- The State University of New
  Jersey, Department of Physics and Astronomy, Piscataway, NJ, USA} 

\author{K.~Reil}
\affiliation{University of Utah, Department of Physics, Salt Lake
  City, UT, USA}

\author{R.~Riehle}
\affiliation{University of Utah, Department of Physics, Salt Lake
  City, UT, USA}

\author{M.~Roberts}
\affiliation{University of New Mexico, Department of Physics and
  Astronomy, Albuquerque, NM, USA }

\author{D.~Rodriguez}
\affiliation{University of Utah, Department of Physics, Salt Lake
  City, UT, USA}

\author{N.~Sasaki}
\affiliation{University of Tokyo, Institute for Cosmic Ray Research,
  Kashiwa, Japan}

\author{S.R.~Schnetzer} 
\affiliation{Rutgers University --- The State University of New
  Jersey, Department of Physics and Astronomy, Piscataway, NJ, USA}

\author{L.M.~Scott}
\affiliation{Rutgers University --- The State University of New
  Jersey, Department of Physics and Astronomy, Piscataway, NJ, USA}

\author{G.~Sinnis}
\affiliation{Los Alamos National Laboratory, Los Alamos, NM, USA}

\author{J.D.~Smith}
\affiliation{University of Utah, Department of Physics, Salt Lake
  City, UT, USA}

\author{P.~Sokolsky}
\affiliation{University of Utah, Department of Physics, Salt Lake
  City, UT, USA}

\author{C.~Song}
\affiliation{Columbia University, Department of Physics and Nevis
  Laboratory, New York, New York, USA}

\author{R.W.~Springer}
\affiliation{University of Utah, Department of Physics, Salt Lake
  City, UT, USA}

\author{B.T.~Stokes}
\affiliation{University of Utah, Department of Physics, Salt Lake
  City, UT, USA}

\author{S.~Stratton}
\affiliation{Rutgers University --- The State University of New
  Jersey, Department of Physics and Astronomy, Piscataway, NJ, USA} 

\author{S.B.~Thomas}
\affiliation{University of Utah, Department of Physics, Salt Lake
  City, UT, USA}

\author{J.R.~Thomas}
\affiliation{University of Utah, Department of Physics, Salt Lake
  City, UT, USA}

\author{G.B.~Thomson}
\affiliation{University of Utah, Department of Physics, Salt Lake
  City, UT, USA}
\affiliation{Rutgers University --- The State University of New
  Jersey, Department of Physics and Astronomy, Piscataway, NJ, USA}

\author{D.~Tupa}
\affiliation{Los Alamos National Laboratory, Los Alamos, NM, USA}


\author{A.~Zech}
\affiliation{Rutgers University --- The State University of New
  Jersey, Department of Physics and Astronomy, Piscataway, NJ, USA}

\author{X.~Zhang}
\affiliation{Columbia University, Department of Physics and Nevis
  Laboratory, New York, New York, USA}

\collaboration{The High Resolution Fly's Eye Collaboration}

\begin{abstract}
We report studies of ultra-high energy cosmic ray composition via analysis of depth of airshower maximum ($\xmax$), for airshower events collected by the High Resolution Fly's Eye (HiRes) observatory. The HiRes data are consistent with a constant elongation rate $d\xmean/d(log(E))$ of $47.9 \pm 6.0 ({\rm stat.}) \pm 3.2 ({\rm syst.})$~g/cm$^2$/decade for energies between 1.6~EeV and 63~EeV, and are consistent with a predominantly protonic composition of cosmic rays when interpreted via the QGSJET01 and QGSJET-II high-energy hadronic interaction models. These measurements constrain models in which the galactic-to-extragalactic transition is the cause of the energy spectrum ``ankle'' at $4 \times 10^{18}$~eV.
\end{abstract}


\maketitle

The observation of a break in the cosmic ray energy spectrum at approximately $6 \times 10^{19}$~eV~\citep{abbasi-prl-100-101101-2008,abbasi-app-32-53-2010,abraham-prl-101-061101-2008} provides evidence that the highest energy cosmic rays are both extragalactic and protonic~\cite{greisen-prl-16-748-1966,zatsepin-jetpl-4-78-1966}. Direct evidence for a proton-dominated composition from airshower data would lend further support to this model, as would the observation of a transition from (heavy) galactic to (light) extragalactic cosmic rays at lower energies. A second feature, the ``ankle'' of the energy spectrum at $4 \times 10^{18}$~eV may be indicative of this transition or it may further strengthen the model in which the end of the cosmic ray spectrum is shaped by interactions with the cosmic microwave background~\cite{berezinsky-icrc-2009}. Composition studies can provide decisive evidence in the choice between interpretations. 

An important clue to chemical composition which is accessible to air fluorescence observatories is the depth of shower maximum $\xmax$ of cosmic ray induced extensive airshowers. Simple arguments~\cite{heitler-qtr-1954,matthews-app-22-387-2005} show that the average value of airshower maximum $\xmean$ will depend logarithmically on the primary energy and atomic mass, and that the {\em elongation rate} $d\xmean/d\loge$ will be constant for unchanging primary compositions. Further, to first order, a nucleus-induced shower may be thought of as a superposition of showers induced by single nucleons. Therefore due to averaging effects we also expect the width of the $\xmax$ distribution at a given energy to be sensitive to the atomic mass of the primary.

The two fluorescence observatories of the High-Resolution Fly's Eye collected data in stereoscopic mode between December 1999 and April 2006. Located on the U.S. Army Dugway Proving Ground in Utah, at a mean elevation of 1,575~m MSL, a mean latitude of $40.16^{\circ}$~N, and separated by 12.6~km, the observatories operated on clear moonless nights. Each detector consisted of an array of telescope modules, each module included a mirror of 3.7~m$^2$ effective area which focused UV light from airshowers on a $16 \times 16$ photomultiplier tube (PMT) camera. The field of view of each PMT subtended a one degree diameter cone of the sky. The HiRes-I detector covered nearly $360^{\circ}$ in azimuth, $3^{\circ}$--$17^{\circ}$ in elevation and was read out by means of sample-and-hold electronics, while the HiRes-II detector covered $3^{\circ}$--$31^{\circ}$ in elevation and was read out by a custom FADC system~\cite{boyer-nima-482-457-2002}. 

The calibration of the HiRes telescope modules has been described previously \cite{abbasi-app-23-157-2005}.  A portable Xenon flash lamp ($\sim{0.5}$\% stability) was used to illuminate each mirror monthly. Between Xenon runs, nightly calibrations were performed using YAG laser light delivered to the cameras via optical fiber~\cite{girard-nima-460-278-2001}. A pulsed nitrogen laser was fired into the atmosphere from various locations within 3.5~km of the two detector sites. An overall accuracy of $\sim{10}$\% RMS is achieved in the HiRes photometric scale. 

Steerable lasers fired patterns of shots which covered the aperture of the HiRes fluorescence detectors, in order to monitor UV attenuation in the atmosphere. The vertical aerosol optical depth (VAOD) was measured to be $0.04 ({\rm mean}) \pm 0.02 ({\rm RMS})$~\cite{abbasi-app-25-74-2006,abbasi-app-25-93-2006}, corresponding to a mean correction of $\sim$15\% upward in energy for an event 25~km distant from the observatory. In the present analysis, the steerable laser measurements were used to compile an hourly database of the atmospheric parameters.

A mirror trigger was initiated if a sufficient number of PMTs were in temporal and spatial coincidence, then a stereo data set was obtained by the time-matching of HiRes-I and HiRes-II triggers. Geometrical reconstruction of stereo events proceeded by determination of the shower-detector plane from each HiRes site, the intersection of these two planes was taken to be the shower core trajectory. The resolution in the shower zenith angle is $0.6^{\circ}$, and the resolution in $R_\mathrm{p}$ (distance of closest approach to the detector) is 1.2\%.

Hit information from multiple tubes in the HiRes-II data only are sorted into discrete time bins. In each bin, FADC signals are converted to a number of photoelectrons $N_\mathrm{pe}$, then adjusted for the effective area of each bin as determined by ray tracing. The geometry of the shower and the atmospheric databases are used to determine the atmospheric slant depth $X$ for each shower bin. Shower segments that have emission angles within 5$^\circ$ of a bin's pointing direction are flagged and not used for fitting due to excessive Cherenkov light contamination. The $N_\mathrm{pe}$ profile is then converted to a profile of the fluorescence light at the shower by a routine that simulates the light production and propagation through the atmosphere, and subtracts the Cherenkov contribution. 

The intensity of fluorescence light emitted from an airshower is proportional to the total ionization energy deposited by the charged particles in the shower~\cite{belz-app-25-57-2006}. We use the average fluorescence yield of several groups~\cite{kakimoto-nima-372-527-1996,nagano-app-20-293-2003,belz-app-25-129-2006} and the spectral distribution given by Ref.~\cite{bunner-phd-1966}, along with the average $dE/dX$ determined from CORSIKA simulations~\cite{heck-karlsruhe-1998} to determine the number of charged particles in the shower as a function of slant depth. 

The shower profile is then fit to a Gaussian function of the age parameter $s(X) = 3X/(X + 2X_\mathrm{max})$ in order to determine the airshower energy and $\xmax$. (Alternatively fitting by the Gaisser-Hillas parametrization~\cite{gaisser-icrc-1977} had little overall effect on the analyses and conclusions presented here.) Further details of the reconstruction used in this analysis are contained in Ref.~\cite{hanlon-phd-2008}.

Ultra-high energy cosmic ray composition studies require a detailed comparison of data with the predictions of airshower models and are hence model dependent. In practice, these models are airshower Monte Carlo simulations. In order to completely understand the effects of the geometrical aperture of the detector, HiRes applies a full detector simulation to the Monte Carlo events. 

HiRes uses libraries of simulated proton- and iron-induced airshowers generated by the CORSIKA~6.003 (6.501)~\citep{heck-karlsruhe-2001} package, using the QGSJET01~\cite{kalmykov-phys_atom_nucl-56-346-1993}, QGSJET-II~\cite{ostapchenko-nucl_phys_proc_suppl-151b-143-2006}, and SIBYLL~\cite{fletcher-prd-50-5710-1994,engel-icrc-1999} hadronic interaction models and the EGS4~\cite{nelson-slac-1985} electromagnetic interaction driver. The number of particles versus slant depth is recorded at 205 points along each shower. 

Detector simulation proceeds by drawing an event from the shower library, assigning it a random core location, zenith and azimuthal angle and determining if it can trigger the detector. The number of charged particles at many discrete points along the shower is determined, and fluorescence light is then propagated from the shower to the detector. Light attenuation by the atmosphere is realistically simulated by using an hourly database describing the measured aerosol content, temperature, and pressure. Ray tracing is performed to determine which phototubes see the light, allowing for photocathode response and inactive space between PMTs. An electronics simulation then determines the pulse height and time for each tube for HiRes-I and forms a FADC waveform for HiRes-II. For all tubes the channel gains, DAC settings, thresholds and channel variances are simulated by using hourly database information. The same trigger algorithms used in hardware are simulated, and if either detector would have been triggered the simulated data is written to disk in the identical format as real data, allowing the study of Monte Carlo events by the same analysis chain. The Monte Carlo set for each hadronic interaction model contains approximately 20~times the number of reconstructed stereo events as the data.

The major challenge in studying cosmic ray composition by the $\xmean$ technique lies in understanding the systematic biases that occur during reconstruction and event selection. Low-energy showers which are nearby the detector may reach $\xmax$ above the field of view of the mirrors, thus biasing a data set towards deeper showers. High-energy showers with small zenith angles are likely to reach maximum below the field of view of the mirrors, resulting in a bias towards shallow showers. 

It is useful to divide the types of biases which can occur into two types, called ``acceptance biases'' and ``reconstruction biases''. Acceptance biases are due to events which fail reconstruction altogether, including detector triggering and event selection effects. Reconstruction biases are due to events which are successfully reconstructed, but with the wrong $\xmax$. The strategy in the following analysis is to choose event selection cuts which minimize the reconstruction bias, and make the acceptance bias as independent of energy as possible. 

After geometrical reconstruction and fitting of the shower profiles to obtain energy and $\xmax$, a set of cuts are applied in order to select an appropriate event sample. The chance probability that the event is due to noise must be less than 1\%, and the $\chi^2/{\rm DOF}$ of the fit must be less than 4. Data must have been taken in good weather conditions. Fit uncertainty in the zenith angle must be less than $2^{\circ}$, the fit uncertainty in $\xmax$ must be less than 40~g/cm$^2$, and the angular RMS with respect to the event plane of hit PMTs must be greater than $0.15^{\circ}$. The zenith angle of the event itself must be less than $70^{\circ}$, and the $R_\mathrm{p}$ with respect to HiRes-II must be at least 10~km. Events are required to have $\xmax$ bracketed by the HiRes-II field of view, and have a shower-detector plane angle between $40^{\circ}$ and $130^{\circ}$. Finally, events with energies $18.2 < \log(E({\rm eV})) < 19.8$ are selected for this analysis. A total of 815 events pass all cuts.

\begin{figure}[t]
\vspace{-0.7cm}
\includegraphics[width=1.0\textwidth]{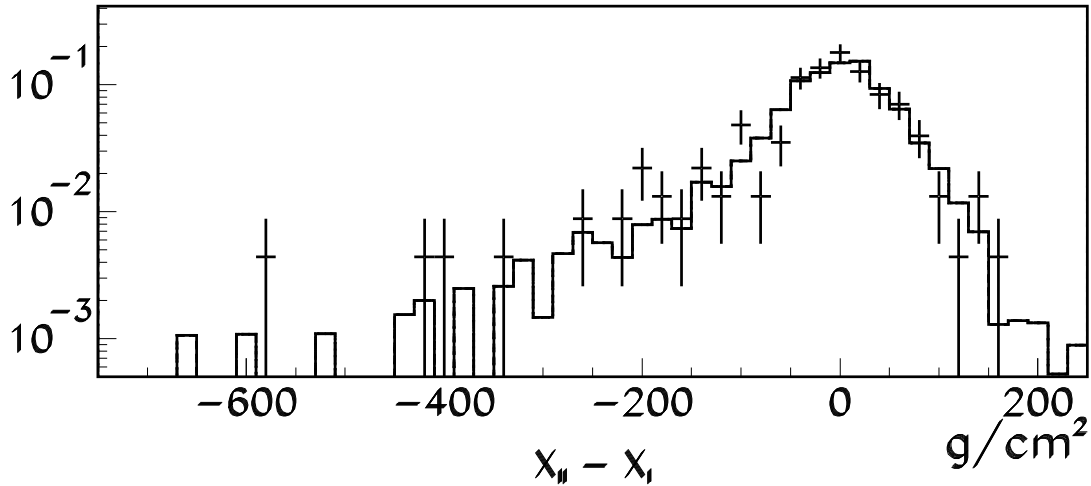}
\vspace{-0.9cm}
\caption{\label{fig:reso_vse_prot_q2_ga} Difference between HiRes-II ($X_{II}$) and  HiRes-I ($X_{I}$) $\xmax$ for HiRes stereo data (points) overlaid with QGSJET-II proton Monte Carlo. $\xmax$ bracketing by HiRes-I is required.}
\end{figure}

In Fig.~\ref{fig:reso_vse_prot_q2_ga} the resolution in $\xmax$ of data and Monte Carlo events are compared by plotting the difference between $\xmax$ as measured by HiRes-I and HiRes-II. The agreement is excellent, including the asymmetry caused by HiRes-I covering only half the range in elevation angle. This supports the use of Monte Carlo to determine $\xmax$ resolution, which is found to be better than 25~g/cm$^2$ over most of the HiRes energy range.

\begin{figure}[h]
\vspace{-0.7cm}
\includegraphics[width=1.0\textwidth]{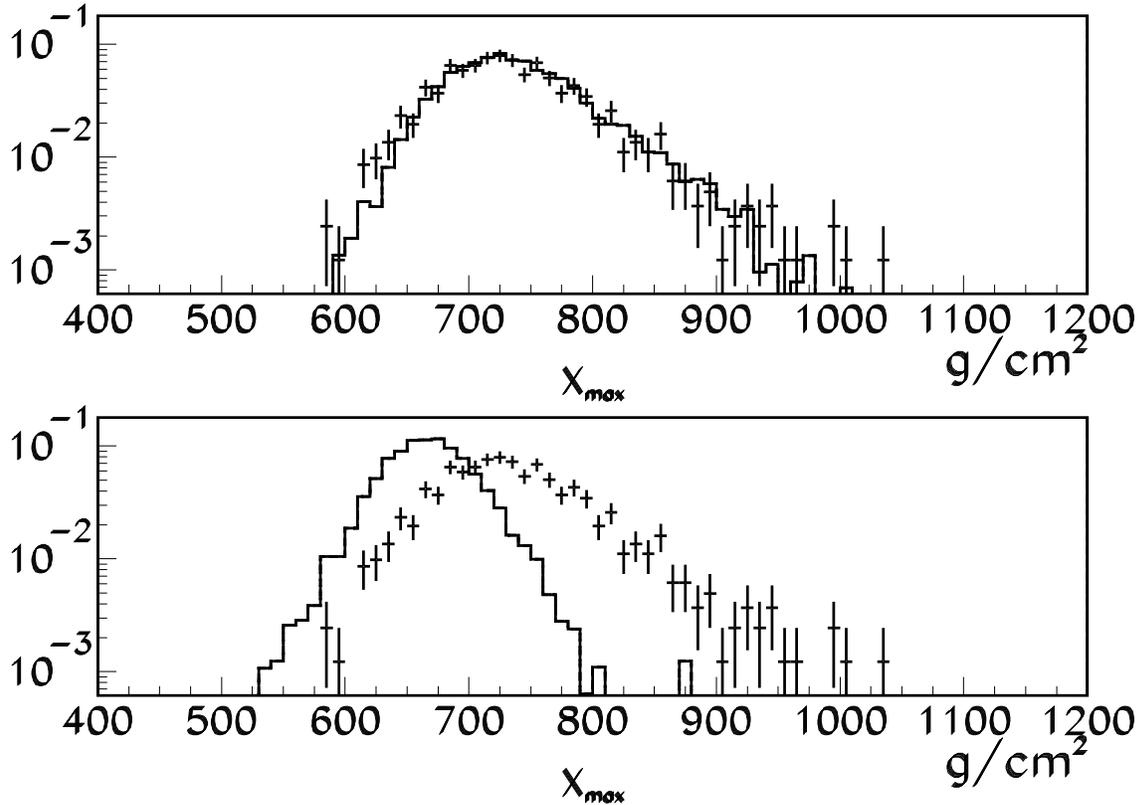}
\vspace{-0.9cm}
\caption{\label{fig:ovlay_xmax} {\em Top:} $\xmax$ overlay of HiRes data (points) with QGSJET-II proton Monte Carlo airshowers after full detector simulation. {\em Bottom:} $\xmax$ overlay of HiRes data (points) with QGSJET-II iron Monte Carlo airshowers.}
\end{figure}

\begin{figure}[h]
  \vspace{-0.7cm}
  \includegraphics[width=1.0\textwidth]{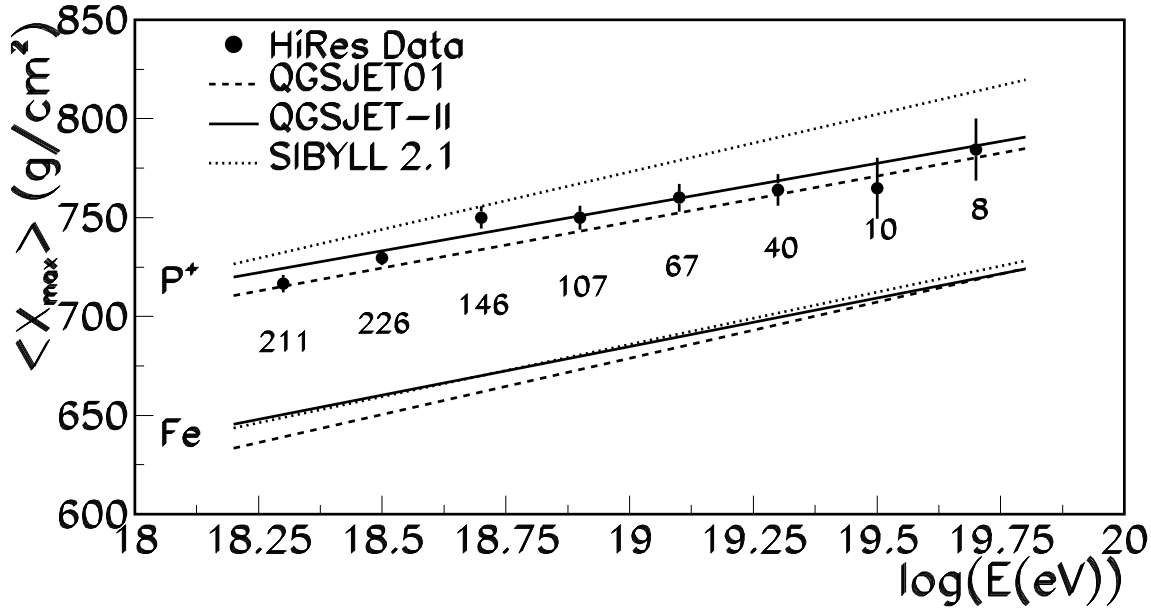}
  \vspace{-0.9cm}
  \caption{\label{fig:nocorr_all_prl} HiRes stereo $\xmean$ compared with the predictions for QGSJET01, QGSJET-II and SIBYLL protons and iron after full detector simulation. The number of events in each energy bin is displayed below the data point.} 
\end{figure}

After application of the cuts above, we compare the distribution in $\xmax$ to the predictions obtained from simulated proton and iron showers. Fig.~\ref{fig:ovlay_xmax} shows the excellent overall agreement between the HiRes stereo data and the predictions of the QGSJET-II proton Monte Carlo. In Fig.~\ref{fig:nocorr_all_prl}, we compare the development with energy of the mean $\xmax$ in HiRes data with the predictions (after full detector simulation) for QGSJET01, QGSJET-II and SIBYLL proton and iron. The data agrees best with the QGSJET-II proton prediction, with a $\chi^2 = 6.9/8$~df. In a linear fit with $\chi^2 = 5.2/(6~{\rm df})$ we measure $47.9 \pm 6.0 ({\rm stat.})$~g/cm$^2$/decade for the elongation rate. Systematic effects are considered below. 

Due to detector, reconstruction, and event selection acceptance effects, the proton and iron ``rails'' in Fig~\ref{fig:nocorr_all_prl} are shifted relative to the raw CORSIKA predictions. However, we find that the shift in mean $\xmax$ for QGSJET01 and QGSJET-II protons is independent of energy to approximately 1.8~g/cm$^2$/decade. We assign a systematic uncertainty of 2.7~g/cm$^2$/decade to the elongation rate based on small variations of event selection cuts. Uncertainties in the energy do not have a large effect on elongation rate results due to the logarithmic energy scale. The choice of VAOD was the main systematic in a previous elongation rate analysis~\citep{abbasi-apj-622-910-2005}, however the use of an hourly atmospheric database in the present analysis renders this source of systematics negligible.

The phototube pointing directions have been confirmed by studies using stars~\citep{sadowski-app-18-237-2002} to within $0.3^{\circ}$, corresponding to a shift in $\xmax$ of approximately 15~g/cm$^2$. Averaging over mirrors, this contributes a net uncertainty of 3.3~g/cm$^2$ to the value of $\xmean$. The subtraction of the Cherenkov light from the phototube signal can introduce an uncertainty in $\xmax$ due to uncertainties in electron multiple scattering. Previous studies~\cite{abbasi-apj-622-910-2005} in which the width of the Cherenkov beam was varied by $2^{\circ}$~(1~$\sigma$) indicated negligible effect on the elongation rate or absolute value of $\xmean$. Finally, a systematic uncertainty of 0.7~g/cm$^2$ is assigned to the absolute value of $\xmean$ in the predictions due to Monte Carlo statistics.

The fluctuations of $\xmax$ as a function of energy are also a probe of primary particle composition. Because the distributions tend to be both asymmetric and possess non-Gaussian tails, care must be taken to use a suitable definition of the $\xmax$ width. The uncorrected RMS and sample standard deviations are biased estimators of the width~\cite{kenney-mos2-1951} and tend to be subject to large fluctuations in distributions with broad tails. 

\begin{figure}
\vspace{-0.7cm}
\includegraphics[width=1.0\textwidth]{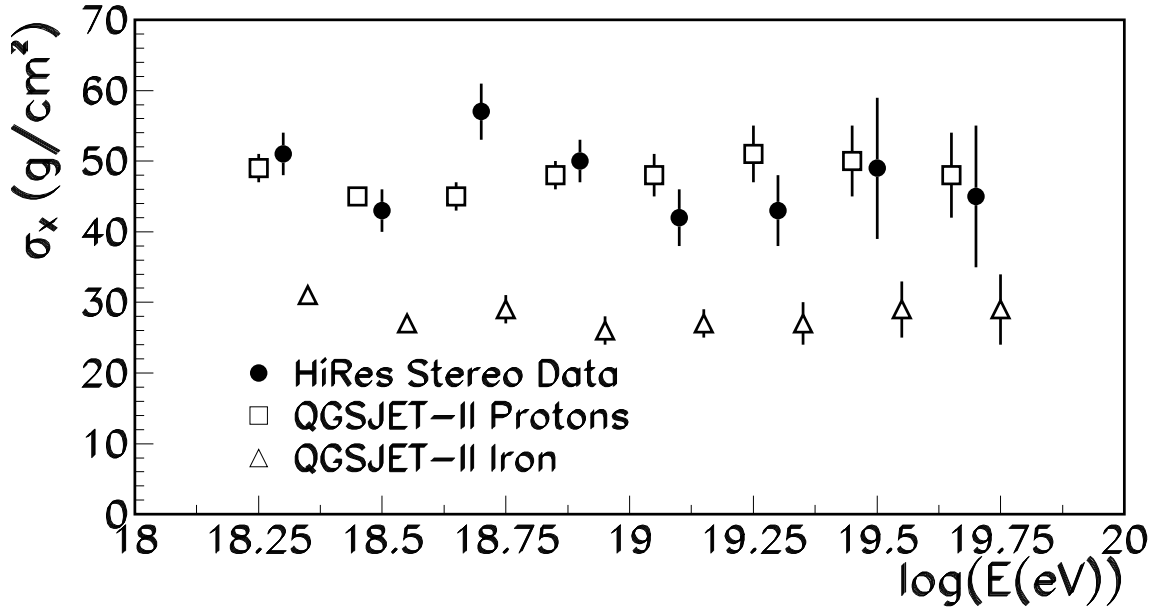}
\vspace{-0.9cm}
\caption{\label{fig:twosig_q2_ga} Results of fitting HiRes stereo data $\xmax$ distribution to Gaussian truncated at $2 \times {\rm RMS}$ (black points). Superimposed are expectations based on QGSJET-II proton (squares) and iron (triangles) Monte Carlo. Monte Carlo points are shown with small offsets in energy to provide separation.}
\end{figure}

In order to focus attention on the center of the $\xmax$ distribution and reduce sensitivity to fluctuations in the tails, the width is quantified as the width $\sigma_X$ of a unbinned likelihood fit to a Gaussian of a distribution truncated at $2 \times {\rm RMS}$. The results of this analysis applied to both the HiRes data and to QGSJET-II proton and iron Monte Carlo are shown in Fig.~\ref{fig:twosig_q2_ga}. The HiRes $\xmax$ width data are consistent with the predictions of QGSJET-II protons. The width of the $\xmax$ distribution of protons within the QGSJET01 model tends to be about 5~g/cm$^2$ broader than that of QGSJET-II, while SIBYLL protons are 2-3~g/cm$^2$ narrower than those of QGSJET-II. Both of these shifts are small compared with statistical uncertainties, particularly at the highest energies. 

In summary, the HiRes data are consistent with a constant elongation rate of $47.9 \pm 6.0 ({\rm stat.}) \pm 3.2 ({\rm syst.})$~g/cm$^2$/decade above 1.6~EeV, and thus with an unchanging composition across the ankle. This places strong constraints on models in which the ankle is the result of a transition from heavy galactic to light extragalactic cosmic rays.

Of the hadronic interaction models tested, the best agreement is with the QGSJET-II pure proton model. Within current uncertainties the data are completely consistent with this model, and close to QGSJET01 pure protons. Comparison with SIBYLL suggests a mixture dominated by light elements. 
The observed constant elongation rate would imply that this mixture is unchanging, or at most steadily changing over nearly two orders of magnitude spanning the energy spectrum ankle. 

The present analysis, taken together with the HiRes spectral results~\cite{abbasi-prl-100-101101-2008,abbasi-app-32-53-2010} on the shape and location of the high-energy cutoff and ankle, suggests the simple picture in which cosmic rays above 1~EeV are protons of extragalactic origin and the end of the energy spectrum is shaped by interactions with the cosmic microwave background.

PHY-9904048, PHY-9974537, PHY-0098826, PHY-0140688, PHY-0245428,
PHY-0305516, PHY-0307098, and by the DOE grant FG03-92ER40732.We
gratefully acknowledge the contributions from the technical staffs of
our home institutions. The cooperation of Colonels E.~Fischer,
G.~Harter and G.~Olsen, the US Army, and the Dugway Proving Ground
staff is greatly appreciated.

{\em Note added.} --- In a recent paper~\cite{abraham-prl-104-091101-2010}, the Pierre Auger collaboration draws different conclusions as to the composition of the highest energy cosmic rays. 


\end{document}